\documentclass[review,number,sort&compress]{elsarticle}

\usepackage{lineno}

\usepackage[english]{babel}
\usepackage[utf8]{inputenc}
\usepackage[sc]{mathpazo}
\usepackage{amsmath}
\usepackage{amssymb}
\usepackage{url}
\usepackage{xspace}
\usepackage{graphicx}
\usepackage{lmodern}

\renewcommand{\vec}[1]{\boldsymbol{#1}}

\newcommand{\code}[1]{\texttt{#1}\xspace}
\newcommand{\eq}[1]{Eq.\,(\ref{eq:#1})\xspace}
\newcommand{\fg}[1]{Fig.\,(\ref{fig:#1})\xspace}
\newcommand{\lsq}{Q}
\newcommand{\ml}{\ln\!{L}}

\DeclareMathOperator*{\ex}{E}
\DeclareMathOperator*{\var}{Var}
\DeclareMathOperator*{\Tr}{Tr}

\newcommand{\phat}{\hat{\vec{p}}}

\title{Application of the Iterated Weighted Least-Squares Fit to counting experiments}

\author[1]{Hans Dembinski\corref{cor1}}
\ead{hdembins@mpi-hd.mpg.de}
\author[1]{Michael Schmelling}
\author[2]{Roland Waldi}
\address[1]{Max Planck Institute for Nuclear Physics, Heidelberg, Germany}
\address[2]{Rostock University, Rostock, Germany}

\cortext[cor1]{Corresponding author}

\begin{document}

\begin{abstract}
Least-squares fits are popular in many data analysis applications, and so we review some theoretical results in regard to the optimality of this fit method. It is well-known that common variants of the least-squares fit applied to Poisson-distributed data produce biased estimates, but it is not well-known that the bias can be overcome by iterating an appropriately weighted least-squares fit. We prove that the iterated fit converges to the maximum-likelihood estimate. Using toy experiments, we show that the iterated weighted least-squares method converges faster than the equivalent maximum-likelihood method when the statistical model is a linear function of the parameters and it does not require problem-specific starting values. Both can be a practical advantage. The equivalence of both methods also holds for binomially distributed data. We further show that the unbinned maximum-likelihood method can be derived as a limiting case of the iterated least-squares fit when the bin width goes to zero, which demonstrates the deep connection between the two methods.
\end{abstract}

\maketitle

\section{Introduction}

In this paper, we review some theoretical results on least-squares methods, in particular, when they yield optimal estimates. We show how they can be applied to counting experiments without sacrificing optimality. The insights discussed here are known in the statistics community~\cite{nw72,cfy76}, but less so in the high-energy physics community. Standard text books on statistical methods and papers, see {\it e.g.} \cite{james2006statistical,baker1984clarification}, correctly warn about biased results when standard variants of the least-squares fit are applied to counting experiments with small numbers of events, but do not show that these can be overcome. The results presented here are of practical relevance for fits of linear models, where the iterated weighted least-squares method discussed in this paper converges faster than the standard maximum-likelihood method and does not require starting values near the optimum.

The least-squares fit is a popular tool of statistical inference. It can be applied in situations with $k$ measurements $\{y_i|i=1, \dots, k\}$, described by a model with $m$ parameters $\vec p = (p_j|j=1,\dots, m)$ that predicts the expectation values $\ex[y_i] = \mu_i(\vec p)$ for the measurements. The measurements differ from the expectation values by unknown residuals $\epsilon_i$ = $y_i - \mu_i(\vec p)$. The solution $\phat$ that minimizes the sum $\lsq(\vec p)$ of squared residuals,
\begin{equation}
\lsq(\vec p) = \sum_{i=1}^{k} \big(y_i - \mu_i(\vec p)\big)^2,
\label{eq:ls}
\end{equation}
is taken as the best fit of the model to the data.

More generally, the measurements and the model predictions can be regarded as $k$-dimensional vectors $\vec y = (y_i|i = 1, \dots, k)$ and $\vec \mu = (\mu_i|i = 1, \dots, k)$, for which one wants to minimize a distance measure.
%A metric space is a vector space in which such distances are defined. In \eq{ls}, we minimize one type of metric, the squared Euclidean distance.
In \eq{ls}, we minimize the squared Euclidean distance. A generalization is the bilinear form
\begin{equation}
\lsq(\vec p) = (\vec{y} - \vec \mu)^T \vec{W} (\vec{y} - \vec \mu),
\label{eq:vls}
\end{equation}
where $\vec{W}$ is a positive-definite symmetric matrix of weights. This variant is called \emph{weighted least squares} (WLS). \eq{ls} is recovered with $\vec{W} = \vec{1}$. An important special case is when the weight matrix is equal to the inverse of the true covariance matrix $\vec{C}$ of the measurements, $\vec{W} = \vec{C}^{-1}$ with $\vec{C} = \ex[\vec y \vec y^T] - \ex[\vec{y}] \ex[\vec{y}]^T$. For uncorrelated measurements, \eq{vls} simplifies to the familiar form
\begin{equation}
\lsq(\vec p) = \sum_{i=1}^{k} \big(y_i - \mu_i(\vec p)\big)^2 \big/ \sigma_i^2,
\label{eq:wls}
\end{equation}
with variances $\sigma_i^2 = \ex[y_i^2] - \ex[y_i]^2$.

Aitken~\cite{aitken1934least} showed in a generalization to the Gauss-Markov theorem~\cite[p.\ 152]{james2006statistical} that minimizing $Q(\vec p)$ with $\vec{W} \propto \vec{C}^{-1}$ produces an optimal, in the sense as detailed below, solution for linear models $\vec{\mu}(\vec p) = \vec{X} \vec p$, where $\vec{X}$ is a constant $k\times m$ matrix. The theorem applies when the covariance matrix $\vec{C}$ is finite and non-singular. Then, $\lsq(\vec p)$ has a unique minimum at
\begin{equation}
\phat = (\vec X^T \, \vec C^{-1} \,\vec X)^{-1} \, \vec X^T \, \vec C^{-1} \, \vec y.
\label{eq:wls_linear_solution}
\end{equation}
The best fit parameters $\phat$ in this case are a linear function of the measurements $\vec y$ with the covariance matrix
\begin{equation}
   \vec C_{p} = (\vec X^T \, \vec C^{-1} \,\vec X)^{-1} .
\label{eq:wls_covmat}
\end{equation}
If the measurements are unbiased, $\ex[\vec y] = \vec \mu$, this solution is the \emph{best linear unbiased estimator} (BLUE). Like all linear estimators, \eq{wls_linear_solution} is unbiased if the input is unbiased. In addition, it has minimal variance of all linear estimators. This is true for any shape of the data distribution and any sample size. These excellent properties may be compromised in practical applications, since the covariance matrix $\vec C$ is often only approximately known.
% If the model $\vec{\mu}(\vec{p})$ is non-linear but analytical, the solution $\phat$ still has minimal variance in the limit of large samples.

The least-squares approach is often regarded as a special case of the more general \emph{maximum-likelihood} (ML) approach. The ML principle states that the best fit of a model should maximize the likelihood $L$, which is proportional to the joint probability of all measurements under the model.
%It is more convenient to work with the logarithm of $L$, and so one maximizes the sum
In practice, it is more convenient to work with $\ml$ rather than $L$, so that the product of probabilities turns into a sum of log-probabilities,
\begin{equation}
\ml(\vec p) = \ln \prod_{i=1}^k P_i(y_i; \vec{p}) = \sum_{i=1}^k \ln P_i(y_i; \vec{p}).
\label{eq:ml}
\end{equation}
Here $P_i(y_i; \vec p)$ is the value of the probability density at $y_i$ for continuous outcomes or the actual probability for discrete outcomes. The ML method needs a fully specified probability distribution for each measurement, while the WLS method uses only the first two moments.

The parameter vector $\phat$ that maximizes \eq{ml} is called the \emph{maximum-likelihood estimate} (MLE). MLEs have optimal asymptotic properties; asymptotic here means in the limit of infinite samples. They are \emph{consistent} (asymptotically unbiased) and \emph{efficient} (asymptotically attaining minimal variance)~\cite{james2006statistical}. In many practical cases of inference, in particular when data are Poisson-distributed, this method is known to produce good estimates also for finite samples. These properties make the ML fit the recommended tool for the practicioner~\cite{baker1984clarification,james2006statistical}.

The WLS fit can be derived as a special case of a ML fit, if one considers normally distributed measurements $y_i$ with expectations $\mu_i$ and variances $\sigma_i^2$, where each measurement has the probability density function (PDF)
\begin{equation}
P_i(y_i;\mu_i,\sigma_i) = \frac{1}{\sqrt{2\pi \sigma_i^2}} \exp\left(-\frac{(y_i-\mu_i)^2}{2\sigma_i^2}\right).
\end{equation}
For fixed $\sigma_i^2$, we obtain \eq{wls} from \eq{ml},
\begin{align}
\ml(\vec p) = - \sum_{i=1}^k \frac{(y_i - \mu_i(\vec p))^2}{2 \sigma_i^2}  + c \equiv -\frac{1}{2} \lsq(\vec p) + c,
\label{eq:ml_wls_normal}
\end{align}
where the constant term $c$ depends only on the fixed variances $\sigma_i^2$. Constant terms do not affect the location $\phat$ of the maximum of $\ml$ and the minimum of $\lsq$. We will often drop them from equations.

This derivation shows that for Gaussian PDFs a ML and a WLS fit give identical results when the same fixed variances are used, even if they are not the true variances. This does not hold in general, but is relevant in this context. When data are Poisson-distributed and have small counts, common implementations of the WLS fit are biased as we will show in the following section. The bias does not originate from the skewed shape of the Poisson distribution however, but rather from the fact that the weights are either biased or not fixed.

To these standard methods, we add the \emph{iterated weighted least-squares} (IWLS) fit~\cite{nw72}. It yields maximum-likelihood estimates when data are Poisson or binomially distributed with only the probabilities as free parameters~\cite{cfy76}. This extends the strict equivalence between ML and WLS fits to a larger class of problems, an extension which is highly relevant in practice, since counts in histograms are Poisson distributed, and counted fractions are binomially distributed with the denominator considered fixed. The iterations are used to successively update estimates of the variances $\sigma_i^2$, which are kept constant during minimization.

When IWLS and ML fits are equivalent, which one is recommended? We conducted toy experiments where IWLS and ML fits are carried out numerically, as is common in practice. We found similar convergence rates for both methods when the model is non-linear, and a significantly faster convergence for the IWLS fit if the model is linear. This makes the IWLS fit a useful addition to the toolbox.

We have seen how the WLS fit can be derived from the ML fit under certain conditions. Inversely, we will show that the unbinned ML fit can be derived as a limiting case from the IWLS fit under weak conditions. The derivation shows that the two approaches are deeply connected.

% the value $\max L(\vec p)$ of the maximized log-likelihood cannot be used as a statistic for goodness-of-fit tests, while the value $\min S(\vec p)$ can.

% There are several distributions which are determined by one parameter,
% conveniently their expectation value (mean).  It necessarily
% also determines
% the variance of the distribution.
% Examples are the Poisson distribution (for integers) and the
% exponential distribution (for real numbers).

% If we have a set of $k$ measurements of numbers with such an
% underlying distribution,
% and their expectation values
% depend on $m < k$ parameters, these parameters can be determined via
% a Maximum Likelihood fit.  In the following sections, three cases
% will be shown where this fit can be performed as an iterated
% minimum chisquare fit, where the variances (\ie errors)
% are kept constant
% in each iteration.

\section{Least-Squares Variants In Use}

Standard variants of the WLS fit used in practice produce biased estimates when the fit is applied to Poisson-distributed data with small counts. The bias is often attributed to the breakdown of the normal approximation to the Poisson distribution, but it is actually related to how the unknown true variances $\sigma_i^2$ in \eq{wls} are replaced by estimates.

We demonstrate this along a simple example. We fit the single parameter $\mu$ of the Poisson-distribution
\begin{equation}
P(n;\mu) = e^{-\mu} \, \mu^{n} / n!,
\label{eq:poisson}
\end{equation}
to $k$ counts $\{n_i|i=1,\dots,k\}$ sampled from it. The maximum-likelihood estimate for $\mu$ can be computed analytically by maximizing \eq{ml}. We solve $\partial\ml/\partial\mu \equiv \partial_\mu \ml = 0$ for $\mu$ and obtain the arithmetic average
\begin{equation}
\hat\mu = \frac{1}{k} \sum_{i=1}^k n_i,
\label{eq:poisson_mle}
\end{equation}
which is unbiased and has minimal variance. We will now apply variants of the WLS fit to the same problem, which differ in how they substitute the unknown true variance.

{\bf Variance computed for each sample.} For a single isolated sample, the unbiased estimate of $\mu$ is $\hat\mu_i = n_i$, with variance $\var[n_i] = \mu \simeq \hat\mu_i = n_i$. This is the origin of the well-known $\sqrt{n}$-estimate for the standard deviation of a count $n$. With this variance estimate, we get
\begin{equation}
\lsq(\mu) = \sum_{i=1}^k (n_i - \mu)^2/n_i.
\end{equation}
This form is called Neyman's $\chi^2$ in the statistics literature~\cite{baker1984clarification}. Replacing the true variance $\mu$ by its sample estimate $n_i$ is an application of the \emph{bootstrap principle} discussed by Efron and Tibshirani~\cite{EfroTibs93}. To obtain the minimum, we solve $\partial_\mu \lsq = 0$ for $\mu$ and obtain the harmonic average
\begin{equation}
\frac{1}{\hat\mu} = \frac{1}{k} \sum_{i=1}^{k} \frac{1}{n_i}.
\end{equation}
The solution is biased and breaks down for samples with $n_i=0$. The variance estimates here are constant (they do not vary with $\hat\mu$), but differ from sample to sample. This treatment ignores the fact that the true variance is the same for all samples in this setup.

{\bf Variance computed from model.} Another choice is to directly insert $\var[n_i] = \mu$ in the formula,
\begin{equation}
\lsq(\mu) = \sum_{i=1}^k (n_i - \mu)^2/\mu.
\label{eq:mupar}
\end{equation}
This form, called Pearson's $\chi^2$~\cite{baker1984clarification}, is a conceptual improvement, because $\mu$ is the exact but unknown value of the variance. However, the variance $\mu$ now varies together with the expectation value $\mu$. Solving $\partial_\mu \lsq = 0$ for $\mu$ yields the quadratic average
\begin{equation}
\hat\mu = \sqrt{\frac{1}{k} \sum_{i=1}^k n_i^2}.
\end{equation}
This estimate is also biased, but can handle samples with $n_i=0$. The bias may come at a surprise, since we used the exact value for the variance after all. The failure here can be traced back to the fact that the variance estimates $\sigma_i^2 = \mu$ are not fixed during the minimization. A small positive bias on $\mu$ in \eq{mupar} leads to a second order increase in the numerator, which is overcompensated by a first order increase of the denominator. In other words, the fit tends to increase the variance even at the cost of a small bias in the expectation when given this freedom, because overall it yields a reduction of $\lsq$.

{\bf Constant variance.} Finally, we simply use $\sigma_i^2 = c$, where $c$ is an arbitrary constant,
\begin{equation}
\lsq(\mu) = \sum_{i=1}^k (n_i - \mu)^2/c.
\end{equation}
We solve $\partial_\mu \lsq = 0$ for $\mu$ and obtain the optimal maximum-likelihood estimate \eq{poisson_mle} as the solution; the constant $c$ drops out.

This seems counter-intuitive, since we used a constant for all samples instead of a value close or equal to the true variance. However, this case satisfies all conditions of the Gauss-Markov theorem. The expectation values are trivial linear functions of the parameter $\mu_i = \mu$. The variances $\sigma_i^2$ are all equal and only need to be known up to a global scaling factor, hence any constant $c$ will do.

% A constant matrix $\vec C$ in \eq{vls} is required to obtain an unbiased solution $\phat$ from minimizing $Q(\vec p)$ for linear models. If not all $\partial_{p_k} C_{ij}$ are zero, extra terms appear in \eq{wls_linear_solution} on the right hand side. Since \eq{wls_linear_solution} is an unbiased solution, another solution with additional terms is biased. The matrix $\vec C$ in \eq{wls_linear_solution} does not have to be the true covariance matrix to make $\phat$ unbiased, this identity is only required if $\phat$ should have minimal variance. If the estimate of $\vec C$ is close to the true covariance, we can hope to retain minimal variance for $\phat$. This idea leads us to the iterated weighted least-squares fit.

We learned that keeping the variance estimates constant during minimization is important, but the estimates should in general be as close to the true variances as possible. An iterated fit can satisfy both requirements.

\section{Iterated Weighted Least-Squares}

The iterated (re)weighted least-squares methods (IWLS or IRLS) are well known in statistical regression \cite{nw72},
and can be applied to
fits with $k$ measurements $\{y_i|i=1, \dots, k\}$ described by a model with $m$ parameters $\vec p = (p_j|j=1,\dots, m)$, which predicts the expectations $\ex[y_i] = \mu_i(\vec p)$ and variances $\var[y_i] = \sigma^2_i(\vec p)$ of each measurement.
We will discuss the special application where the $y_i$ are entries of a histogram. One then minimizes the sum of squared residuals
\begin{equation}
\lsq(\vec p) = \sum_{i=1}^{k} (y_i - \mu_i(\vec p)\big)^2/\sigma_i^2(\phat),
\label{eq:iwls}
\end{equation}
where the $\sigma_i^2$ are constant within one iteration of the fit and computed from the model using the parameter estimate $\phat$ that minimized $\lsq(\vec p)$ in the previous iteration. A convenient choice for the first iteration is $\sigma_i^2 = 1$. One iterates until $\phat$ converges.

In particle physics, we often work with samples drawn from two monoparametric distributions of the exponential family:
\begin{itemize}
\item Poisson distribution. Example: fitting a distribution function to a histogram of counts.
\item Binomial distribution with fixed number of trials. Example: fitting an efficiency function to two histograms with generated and accepted events.
% \item \em{SHOULD WE CONSIDER ALSO THE MULTINOMIAL CASE?}
\end{itemize}
Charles, Frome, and Yu~\cite{cfy76} derived that the IWLS fit gives the exact same result as the ML fit for a family of distributions. We demonstrate this in the appendix for the special distributions discussed here.

The Hessian matrices of second derivatives are also equal up to a constant factor, $\partial_{p_l} \partial_{p_m} \lsq = -2 \partial_{p_l} \partial_{p_m} \ml$. The inverse of the Hessian is an estimate of the covariance matrix of the solution, an important uncertainty estimate in practical applications.

We emphasize that the equivalence does not depend on the size of the data sample or on the functional form of the model that predicts the expectation values for the measurements. In particular, when the IWLS fit is applied to histograms, it is not biased by small counts per bin or even empty bins.

\subsection{Including systematic uncertainties}

A formal discussion of how systematic uncertainties can be handled with the IWLS fit is outside of the scope of this paper, but we note that it can include systematic uncertainties. Barlow~\cite{2017arXiv170103701B} discusses how correlated systematic uncertainties can be handled in a least-squares fit. One minimizes \eq{vls} in each iteration with a matrix
\begin{equation}
\vec C = \vec{C}'(\phat) + \vec C_\text{sys}(\phat),
\end{equation}
where $\vec{C}'$ is the current estimate of the stochastic covariance computed from the previous solution, and $\vec C_\text{sys}$ is a current covariance matrix that represents the systematic uncertainties of the measurements. The matrix $\vec C_\text{sys}$ may be a function of the parameter vector. Like the covariance matrix $\vec{C}'$, it is kept constant during each iteration, and updated between iterations using the current value of $\phat$. This approach has been successfully applied in a combination of measurements from the CDF and D0 experiments~\cite{Aaltonen:2013wca}.

\subsection{IWLS or ML fit?}

When the IWLS and the ML fits are equivalent, which one should be used in practice? The two methods produce the same results in analytical problems, but can have different performance in numerical problems. In practice, the extrema of the log-likelihood function $\ml(\vec p)$ and the weighted least-squares function $\lsq(\vec p)$ are usually found with a local optimizer, like the \code{MIGRAD} algorithm in the \code{MINUIT} package~\cite{1975CoPhC..10..343J,iminuit}. Computing the functions is sometimes expensive, when the fitted data sets are large and the model has many parameters. Numerical methods are therefore judged based on the number of function evaluations required to converge to the optimum within some tolerance. Another criterion is robustness, the ability to converge to the right optimum from a point in the neighborhood of the solution.

To address these points, we conducted toy experiments with Poisson-dis\-trib\-ut\-ed counts $n_i$ and find that the ML method requires less function evaluations than the IWLS methods in general. However, the rate of convergence of the IWLS method can be greatly accelerated, when the model that computes the count expectation $E[n_i] = \mu_i(\vec p)$ is linear in the parameters, $\mu_i = \vec{X}_i \, \vec{p}$, where $\vec {X}_i$ is a vector of constants. The maximum of $\ml(\vec p)$ usually cannot be found analytically in this case, but the minimum of $\lsq(\vec p)$ is given by \eq{wls_linear_solution} in each iteration of the IWLS fit. When the computing time is dominated by the evaluation of $\lsq(\vec p)$ or $\ml(\vec p)$, solving the IWLS fit is faster than the ML fit. The IWLS fit also does not require a problem-specific starting point for the optimization in this case. We call this special variant the L-IWLS fit. All three methods are able to handle fits that have bounded parameters, which are common in particle physics. In our toy experiments, the parameters are bounded to be non-negative. Details are given in the next section.

Whether the ML or the IWLS fits are more robust in the above sense is more difficult to say. No general proofs can be given for either method. Our toy studies suggest the following order of increasing robustness: IWLS, L-IWLS, ML. In some toy experiments, the IWLS methods require many more iterations than average, producing a long tail in the distribution of iteration counts. Such tails are not observed for the ML fit. It is likely, however, than a more sophisticated implementation of the IWLS fit than ours could improve the robustness of this method.

\subsection{Performance in toy experiments}

\begin{figure}
\includegraphics[width=\columnwidth]{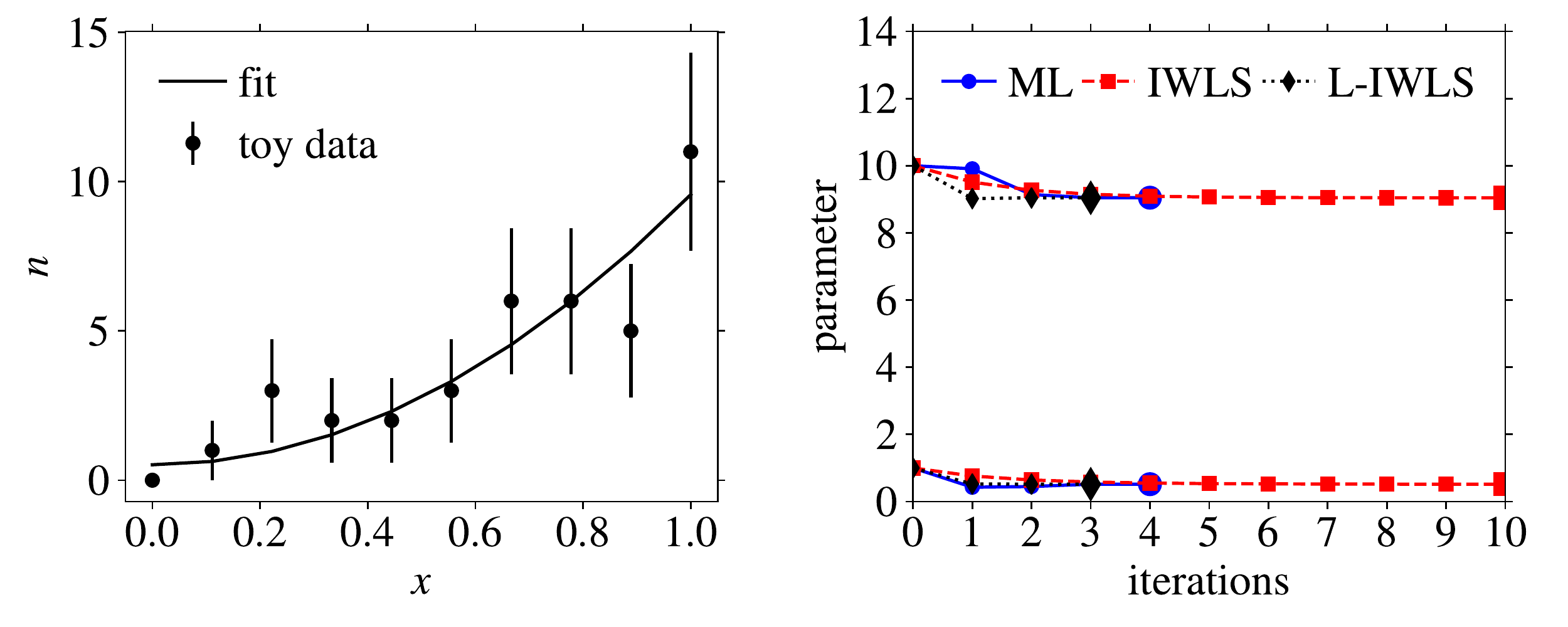}
\caption{Example of a toy data set (points) to test the performance of the ML, IWLS, and L-IWLS fits (see text). \emph{Left:} The curve represent the fit result of the three methods for this data set (which are identical). \emph{Right:} Intermediate parameter states during the optimization for the ML, IWLS, and L-IWLS fits (see text), in each iteration of the respective algorithms until the stopping criterion is reached. The L-IWLS fit often converges quadratically. The IWLS fit is slowed down by the artifical dampening that we introduced to avoid oscillations.}
\label{fig:poisson_example_toy}
\end{figure}

We compare the performances of ML, IWLS, and L-IWLS fits in a series of 1000 toy experiments with Poisson-distributed samples. We use a linear model for the expectation with two parameters, $\mu(x, \vec p) = (p_0 + p_1\, x^2)$, with $x\in [0, 1]$ as an independent variable. For the true parameters $\vec p_\text{truth} = (1, 10)$, we simulate 10 pairs $(x_i, n_i)$. The $x_i$ are evenly spaced over the interval $[0, 1]$, $\mu_i$ is calculated for each $x_i$ based on the true parameters, and finally a random sample $n_i$ is drawn for each $\mu_i$ from the Poisson distribution. The model is then fitted to each toy data set using the following three methods. One of the toy experiments is shown in \fg{poisson_example_toy}.

\begin{itemize}
\item ML fit: Starting from \eq{ml_poisson} we use the \code{MIGRAD} algorithm from the \code{MINUIT} package to find the minimum. We pass the exact analytical gradient to \code{MIGRAD} for this problem, replacing the numerical approximation that \code{MINUIT} uses otherwise. We restrict the parameter range to $p_k \ge 0$ and add an epsilon to $\mu$ whenever it appears in a denominator to avoid division by zero.

\item IWLS fit: We use Newton's method to update $\vec{p}$,
\begin{equation}
\vec{p}_{n+1} = \vec{p}_{n} - \vec{H}^{-1} \, \vec{\partial}_{\vec{p}} \lsq,
\end{equation}
with the exact analytical gradient $\vec{\partial}_{\vec{p}} \lsq$ and Hessian $\vec{H}$ for this problem. Since the model is linear and the function $\lsq$ quadratic, Newton's method yields the exact solution for the given gradient and Hessian matrix, but without taking the boundary condition $p_k \ge 0$ into account. We resolve this in an ad hoc way, by setting negative parameter values are set to zero.

Since the covariance matrix is fixed in each Newton step, each step fulfills the requirements of the IWLS method. We update the covariance matrix after each step for the computation of the next step. To check for convergence, we use the \code{MINUIT} criterion, which is based on the estimated distance-to-minimum and deviations in the diagonal elements of the inverted Hessian~\cite{1975CoPhC..10..343J}.

This approach works very well for most toy experiments, but in some rare cases ($< 1\,\%$) the solution starts to oscillate indefinitely between two states. We resolve this again in an ad hoc way by averaging the updated parameter vector with the previous one, $\vec{p}_{n+1} := (\vec{p}_{n+1} + \vec{p}_{n})/2$ after each Newton step. This slows down the convergence rate drastically, but avoids the oscillations.

\item L-IWLS fit: We solve \eq{wls_linear_solution} with the \code{NNLS}~\cite{doi:10.1137/1.9781611971217} algorithm as implemented in \code{SciPy}~\cite{scipy}, and iterate. It solves \eq{wls_linear_solution} under the boundary condition $p_k \ge 0$. To check for convergence, we again use the \code{MINUIT} criterion.
\end{itemize}

\begin{figure}
\includegraphics[width=\columnwidth]{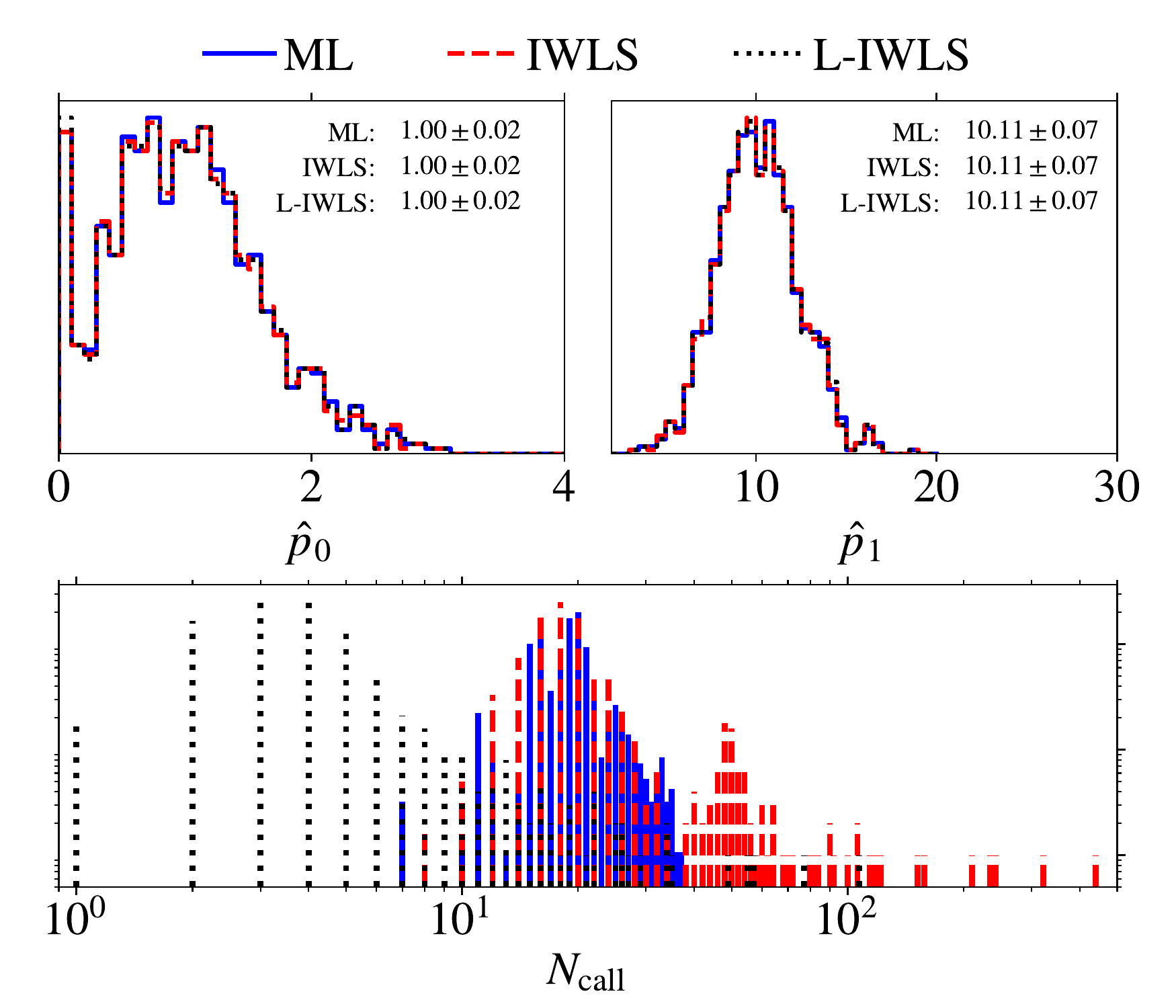}
\caption{Application of the maximum-likelihood (ML), the iterated weighted least-squares (IWLS) fit and its specialization for linear models (L-IWLS) to 1000 toy experiments with Poisson-distributed samples (see text). {\it Top row:} Histograms of the two fitted parameters of the model $\hat p_0$ and $\hat p_1$ are shown, overlayed for all three fit methods. The histograms are nearly identical. {\it Bottom row:} Normalized histograms of the number of evalutions of the model function for the the fit methods in double-logarithmic scale.}
\label{fig:poisson_example}
\end{figure}

We note that our application of the general IWLS fit to a problem with a linear model is artifical. We only do this here to compare all three fitting methods on the same problem. For the IWLS and ML fits, we use the optimistic starting point $\vec p_\text{truth} = (1, 10)$. The ML and IWLS fits therefore run under ideal conditions compared to the L-IWLS fit, which does not require a specific starting point. In practice, one will usually start with a less ideal starting point, which slows down the convergence of ML and IWLS fits compared to the L-IWLS fit.

In case of the ML and IWLS fits, we increase the call counter for each evaluation of $\lsq$ or $\ml$ and each evaluation of their gradients for all values of $z$ by one. In case of the L-IWLS fit, we count one application of the NNLS algorithm as one call, since it requires essentially one computation of the gradient.

The results are shown in \fg{poisson_example}. As expected, the results are equal within the numerical accuracies of the numerical algorithms, which stop when \code{MINUIT}'s standard convergence criterion is reached. This criterion roughly gives a precision of about $10^{-3}$ in the parameter relative to its uncertainty.

The average number of calls required to converge is different: 19.3 for ML, 23.6 for IWLS, and 4.8 for L-IWLS. The L-IWLS fit is the fastest to converge, requiring only a quarter of the function evaluations of the ML fit. The IWLS fit is the slowest, it requires about 20\% more calls on average than the ML fit. This is mainly due to artificial dampening. In cases where the dampening is not needed, the IWLS fit converges as rapidly as the L-IWLS fit. Since we chose a linear model for this performance study in order to compare all three methods, a Newton's step computes the exact solution to the fitting problem for the current covariance matrix estimate.

An investigation shows that the convergence issues of the IWLS fit appear when a parameter of the model is very close to zero. If this is not the case and no dampening is applied, the IWLS and L-IWLS fits produce identical results. \code{MINUIT} was designed to handle such cases well and shows a much more stable convergence rate. This suggests that the issues of the IWLS fit can be overcome as well with a more sophisticated implementation, but this comes at the cost of a slower convergence in favorable cases. The overall performance of the IWLS fit will probably not surpass that of the more straight-forward ML fit.

In conclusion, we recommend the L-IWLS fit for linear models and the ML fit for non-linear models.

\section{Unbinned maximum-likelihood from IWLS}

In the introduction, we reviewed how the WLS fit can be derived as a special case of the ML fit, when measurements are normally distributed with known variance. Alternatively, the WLS fit can also be derived from geometric principles without relying on the ML principle. We will now show that the unbinned ML fit can be derived as a limiting case of the IWLS fit.

For the unbinned ML fit of a known probability density $f(x; \vec p)$ of a continuous stochastic variable $x$ with parameters $\vec{p}$, one maximizes the sum of logarithms of the model density evaluated at the measurements $\{x_i| i = 1 \dots k \}$,
\begin{equation}
\ml(\vec p) = \sum_{i=1}^k \ln f(x_i; \vec p).
\label{eq:ml_unbinned}
\end{equation}
The maximum is found by solving the system of equations $\partial_{p_j} \ml(\vec{p}) = 0$. The density $f(x; \vec p)$ must be at least once differentiable in $\vec p$.

To derive these equations as a limit of the IWLS fit, we assume that $f(x;\vec{p})$ is finite everywhere in $x$, so that the probability density is not concentrated in  discrete points.

We start by considering a histogram of $k$ samples $x_i$. Since the samples are independently drawn from a PDF, the histogram counts $n_l$ are uncorrelated and Poisson-distributed. Following the IWLS approach, we minimize the function
\begin{equation}
\lsq(\vec p) = \sum_{l} \frac{(n_l - k P_l)^2}{k \hat P_l}
\end{equation}
and iterate, where $P_l(\vec p) = \int_{x_l}^{x_l+\Delta x} f(x; \vec p) \,\text{d}x$ is the expected fraction of the samples in bin $l$, and $\hat P_l = P_l(\phat)$ is the value based on the fitted parameters $\phat$ from the previous iteration. Expansion of the squares yields three terms,
\begin{equation}
\lsq(\vec p) = \sum_l \frac{n_l^2}{k \hat P_l} - 2 \sum_l \frac{n_l\,P_l}{\hat P_l} + k \sum_l \frac{P_l^2}{\hat P_l}.
\end{equation}
The first term is proportional to $1/\Delta x$, but not a function of $\vec p$. Therefore it does not contribute to the minimum obtained by solving the equations $\partial_{p_j} \lsq(\vec p) = 0$. We drop it in the following and consider only the second and third term, which both are functions of $\vec p$.

We investigate the limit $\Delta x \rightarrow 0$. Since $f(x)$ is finite everywhere, we have ultimately either zero or one count in each bin. With $P_l \rightarrow f(x_l; \vec p) \, \Delta x$, the second term has a finite limit
\begin{equation}
\sum_l \frac{n_l \, P_l}{\hat P_l}
\xrightarrow{\Delta x \rightarrow 0}
\sum_{i=1}^k \frac{f(x_i; \vec p) \, \Delta x}{f(x_i; \phat) \, \Delta x}
= \sum_{i=1}^k \frac{f(x_i; \vec p)}{f(x_i; \phat)},
\end{equation}
where only bins around the measurements $x_i$ with one entry contribute ($n_l=1$),
and the bin widths cancel. The third term also has a finite limit,
\begin{equation}
\sum_l \frac{P_l^2}{\hat P_l} \xrightarrow{\Delta x \rightarrow 0} \sum_l \frac{f^2(x_l; \vec p)\,(\Delta x)^2}{f(x_l; \phat)\,\Delta x} = \int \frac{f^2(x; \vec p)}{f(x; \phat)}\,\text{d}x.
\end{equation}
One $\Delta x$ cancels in the ratio and in the limit $\Delta x\to0$ the remaining sum is the very definition of a Riemann integral.

We now consider the derivatives $\partial_{p_j}\lsq(\vec p)$ in the limit of many iterations. We assume that the iterations converge, so that the previous solution $\phat$ approaches the next solution $\vec p$. We get
\begin{equation}
\begin{aligned}
\partial_{p_j} \lsq(\vec p) &= -2 \sum_{i=1}^k \frac{\partial_{p_j} f(x_i; \vec p)}{f(x_i; \phat)} + k \int \frac{2 f(x; \vec p) \, \partial_{p_j} f(x; \vec p)}{f(x; \phat)} \, \text{d}x \\
&\xrightarrow{\phat \rightarrow \vec p}
- 2 \sum_{i=1}^k \frac{\partial_{p_j} f(x_i; \vec p)}{f(x_i; \vec p)}
+ 2 k \partial_{p_j} \int f(x; \vec p) \, \text{d}x.
\end{aligned}
\label{eq:lsq_limit}
\end{equation}
The last term vanishes in the limit, because %$\int \partial_{p_j} f(x;\vec p)\,\text{d}x = \partial_{p_j} \int f(x;\vec p)\,\text{d}x = \partial_{p_j} 1 = 0$.
$\int f(x;\vec p)\,\text{d}x = 1$ is constant.

We finally obtain the equivalence
\begin{equation}
\partial_{p_j} \lsq(\vec p)
\xrightarrow{\Delta x
\rightarrow 0,\,\phat \rightarrow \vec p}
- 2\sum_{i=1}^k \frac{\partial_{p_j} f(x_i; \vec p)}{f(x_i; \vec p)} \\
= -2 \sum_{i=1}^k \partial_{p_j} \ln f(x_i; \vec p)
= -2 \partial_{p_j} \ml(\vec p).
\end{equation}
The derivatives are equal up to a constant factor, which means that the solutions of $\partial_{p_j} \lsq(\vec p) = 0$ and $\partial_{p_j} \ml(\vec p) = 0$ are equal. In other words, the IWLS solution in the limit of infinitesimal bins is found by minimizing the negative log-likelihood of the probability density.  The latter is effectively a shortcut to the solution, which does not require iterations.

We showed the equivalence for the case when measurements consist of a single variable $x_i$ per event for simplicity, but it also holds for the general case of a set of n $n$-dimensional vectors $\{\vec x_{i} | i=1 \dots k\}$ with $\vec x_i = (x_{ji} | j=1\dots n)$ and a corresponding $n$-dimensional probability density $f(\vec x; \vec p)$. In this case, one would repeat the derivation starting from an $n$-dimensional histogram.

The derivation provides some insights.
\begin{itemize}
    \item The absolute values of $\lsq(\phat)$ and $-2\ml(\phat)$ at the solution $\phat$ are not equal. They differ by an (infinite) additive constant.
    \item The derivatives $\partial_{p_j} \lsq(\vec p)$ and $-2 \partial_{p_j} \ml(\vec p)$ differ in general when $\vec p$ is not the solution $\phat$, because the second term in \eq{lsq_limit} does not vanish for $\vec p \ne \phat$.
\end{itemize}
In practice, the second point means that the \code{MINUIT} package produces the same error estimates for the solution $\phat$ if the \code{HESSE} algorithm is used, but not if the \code{MINOS} algorithm is used. The \code{HESSE} algorithm numerically computes and inverts the Hessian matrix of second derivatives at the minimum, which gives identical results for $\lsq$ and $-2\ml$. The \code{MINOS} algorithm scans the neighborhood of the minimum, which for $\lsq$ and $-2\ml$ usually has a different shape.

\section{Notes on goodness-of-fit tests}\label{sec:gof}

% work that in:
% If true variances $\sigma_i^2$ are used in the computation, $\lsq(\vec p)$ evaluated at the true parameter values follows Pearson's $\chi^2$-distribution~\cite{doi:10.1080/14786440009463897} with expectation $k$, and $\lsq(\phat)$ evaluated at the estimate $\phat$ follows a $\chi^2$-distribution with expectation $(k-m)$. If the data are not normal-distributed, this is still true in the asymptotic limit. This property is often used to decide whether a particular model matches the data, we will come back to it in section \ref{sec:gof}. In the literature, the weighted least-squares fit is sometimes called \emph{minimum-chisquare} fit due to this property~\cite{berkson1980,baker1984clarification}.

For a goodness-of-fit (GoF) test, one computes a test statistic for a probabilistic model and a set of measurements. The test statistic is designed to have a known probability distribution when the measurements are truly distributed according to the probabilistic model. If the value for a particular model is very improbable, the model may be rejected.

It is well-known that the minimum value $\lsq(\phat)$ is $\chi^2$-distributed with expectation $(k-m)$, if the measurements are normally distributed, where $k$ and $m$ are the number of measurements and number of fitted parameters, respectively~\cite{james2006statistical}. This GoF property is so useful and frequently applied, that the function $\lsq(\vec p)$ is often simply called \emph{chi-square}.

In general, $\lsq(\phat)$ is not $\chi^2$-distributed for measurements that are not normally distributed around the model expectations. Approximately, it holds for Poisson and binomially distributed measurements when counts are not close to zero, and fractions are neither too close to zero or one. Stronger statements can be made about the expectation value of $\lsq(\phat)$. For linear models with $m$ parameters and $k$ unbiased measurements with known covariance matrix $\vec C$, the expectation of $\lsq(\phat)$ is guaranteed to be
\begin{equation}
\ex[\lsq(\phat)] = k - m,
\label{eq:liwls_gof}
\end{equation}
regardless of the sample size and the distribution of the measurements, as shown in the appendix. Therefore, the well-known quality criterion that the \emph{reduced} $\chi^2$ should be close to unity, $\lsq(\phat)/(k-m) \simeq 1$, is often useful even if measurements are not normally distributed.

We saw previously that $-2\ml(\phat)$ differs from $\lsq(\phat)$ by an infinite additive constant, which is a hint that it cannot straight-forwardly replace the latter as a GoF statistic. When used with unbinned data, $\ml(\phat)$ is ill-suited as a GoF test statistic. Heinrich~\cite{2003sppp.conf...52H} presented striking examples when $\ml(\phat)$ carries no information of how well the model fits the measurements. Cousins~\cite{cousins2013,cousins2016} gave an intuitive explanation for this fact. The IWLS fit provides a maximum-likelihood estimate for measurements that follow a Poisson or binomial distribution and a GoF test statistic as a side result, which in general is not exactly $\chi^2$-distributed, but its distribution can often be obtained from a Monte Carlo simulation.

\section{Conclusions}

An iterated weighted least-squares fit applied to measurements, which are Poisson- or binomially distributed around model expectations, provides the exact same solution as a maximum-likelihood fit. This holds for any model and any sample size. When the two fit methods are equivalent, the maximum-likelihood fit is still recommended, except when the model is linear. In this case, the minimum of the weighted least-squares problem can be found analytically in each iteration, which usually needs less computations overall than numerically maximizing the likelihood and requires no problem-specific starting point. The iterated weighted least-squares fit provides a goodness-of-fit statistic in addition, while the maximum-likelihood fit usually does not. Of course, a goodness-of-fit statistic can always be separately computed after the optimization, but in case of the maximum-likelihood it requires implementing two functions in a computer program instead of one.

Whether the two fit methods give equivalent results depends only on the probability distribution of the measurements around the model expectations. Here we presented proofs of the equivalence for Poisson and binomial distributions. In the statistics literature~\cite{nw72,cfy76}, more general proofs are given that hold also for some other distributions.

\section{Acknowledgments}

We thank Bob Cousins for a critical reading of the manuscript and for valuable pointers to the primary statistical literature, and the anonymous reviewer for suggestions to clarify additional points.

\appendix

\section{Equivalence of ML and ILWS for Poisson-distributed data}

A common task is to fit a model to a histogram with $k$ bins, each with a count $n_i$. Especially in multi-dimensional histograms some bins may have few or even zero entries. This poses a problem for a conventional weighted least-squares fit, but not for a ML fit or an IWLS fit.

A ML fit of a model with $m$ parameters $\vec p = (p_j|j=1,\dots, m)$ to a sample of $k$ Poisson-distributed numbers $\{n_i|i=1,\dots, k\}$ with expectation values $\ex[n_i] = \mu_i(\vec p)$ is performed by maximizing the log-likelihood
\begin{equation}
\ml(\vec p) = \sum_{i=1}^k n_i \ln \mu_i - \sum_{i=1}^k \mu_i,
\label{eq:ml_poisson}
\end{equation}
which is obtained by taking the logarithm of the product of Poisson probabilities (\ref{eq:poisson}) of the data under the model, and dropping terms that do not depend on $\vec p$.

To find the maximum, we set the $m$ first derivatives
\begin{equation}
\frac{\partial \ml}{\partial p_j}
= \sum_{i=1}^k \frac{n_i}{\mu_i}
 \frac{\partial \mu_i}{\partial p_j}
 - \sum_{i=1}^k
 \frac{\partial \mu_i}{\partial p_j}
\end{equation}
for $j = 1$ to $m$ to 0. We get a system of equations
\begin{equation}
\sum_{i=1}^k \frac{n_i - \mu_i}{\mu_i}
\, \frac{\partial \mu_i}{\partial p_j} = 0.
\label{eq:ml_poisson_diff}
\end{equation}

We now approach the same problem as an IWLS fit. The sum of weighted squared residuals is
\begin{equation}
\lsq = \sum_{i=1}^k \frac{(n_i - \mu_i)^2}{\hat \mu_i},
\label{eq:wls_poisson}
\end{equation}
where $\hat \mu_i$ is the expected variance computed from the model, using the parameter estimate $\phat$ from the previous iteration. To find the minimum, we again set the $m$ first derivatives to 0 and obtain
\begin{equation}
\frac{\partial \lsq}{\partial p_j} =
-2\sum_{i=1}^k \frac{n_i - \mu_i}{\hat\mu_i}
\, \frac{\partial \mu_i}{\partial p_j} = 0.
\label{eq:wls_poisson_diff}
\end{equation}

\eq{wls_poisson_diff} and \eq{ml_poisson_diff} yield identical solutions in the limit $\hat \mu_i \rightarrow \mu_i$, and so do their solutions. The limit is approached by iterating the fit, so that we actually obtain the maximum-likelihood estimate from the IWLS fit. Remarkably, this does not depend on the size of the counts $n_i$ per bin. The equivalence holds even when many bins with zero entries are present. To obtain this result, the $\hat \mu_i$ must be constant. If $\hat \mu_i$ was replaced by $\mu_i$ in \eq{wls_poisson}, extra non-vanishing terms would appear in \eq{wls_poisson_diff}.

As already mentioned, when the expectations are linear functions, the unique analytical solution to \eq{wls_poisson_diff} is given by \eq{wls_linear_solution}, with $\vec C^{-1} = (\delta_{ij}/\hat \mu_i|i,j = 1,\dots,k)$ and $y_i = n_i$. An analytical solution of \eq{ml_poisson_diff} is not known to the authors. %finding it is made difficult by the non-constant $\mu_i$ in the denominator.
The IWLS fit converges faster than the ML fit in this case.

\section{Equivalence of ML and IWLS for binomially distributed data}

Another common task is to obtain an efficiency function of a selection or trigger as a function of an observable. One collects a histogram of generated events with bin contents $N_i$, and a corresponding histogram of accepted events with bin contents $n_i$. The $N_i$ are considered as constants here, while the $n_i$ are drawn from the binomial distribution. The goal is to obtain a model function that best describes the efficiencies $\epsilon_i$ that best describe the drawn samples $n_i$. A single least-squares fit will give biased results when many $n_i$ are close to either 0 or $N_i$, but not a ML or an IWLS fit.

A ML fit of a model with $m$ parameters $\vec p = (p_j|j=1,\dots, m)$ for a sample of $k$ binomially distributed numbers $\{n_i|i=1,\dots, k\}$ with expectations $\ex[n_i] = \mu_i(\vec p) = \epsilon_i(\vec p) \, N_i$ is performed by maximizing the log-likelihood
\begin{equation}
\ml(\vec p) = \sum_{i=1}^k n_i \ln \mu_i
 + \sum_{i=1}^k (N_i-n_i) \ln(N_i - \mu_i),
\label{eq:ml_binom}
\end{equation}
which is obtained by taking the logarithm of the product of binomial probabilities to observe $n_i$ when $\mu_i = \epsilon_i\,N_i$ are expected,
\begin{equation}
P(n_i;\mu_i,N_i) = \binom{N_i}{n_i} \, \epsilon_i^{n_i} \, (1-\epsilon_i)^{N_i-n_i} = \binom{N_i}{n_i} \frac{\mu_i^{n_i} (N_i-\mu_i)^{N_i-n_i}} {N_i^{N_i}},
\end{equation}
and dropping terms that do not depend on $\vec p$. A binomial distribution has two parameters $(\mu_i,N_i)$, but it is a monoparametric distribution in this context  since the $N_i$ are known and only the $\mu_i$ are free parameters.

Again we set the $m$ first derivatives
\begin{equation}
\frac{\partial \ml}{\partial p_j}
= \sum_{i=1}^k \frac{n_i }{ \mu_i}\,\frac{\partial \mu_i}{\partial p_j}
   - \sum_{i=1}^k \frac{N_i - n_i}{N_i - \mu_i}\,
 \frac{\partial \mu_i}{\partial p_j}
\end{equation}
to zero for $j = 1$ to $m$. The minimum is obtained by solving
\begin{equation}
\sum_{i=1}^k \frac{n_i - \mu_i}{\mu_i (1 - \mu_i/N_i)} \,
 \frac{\partial \mu_i}{\partial p_j} = 0.
\label{eq:ml_binom_diff}
\end{equation}

For the IWLS fit, we need to minimize the sum
\begin{equation}
\lsq(\vec p) = \sum_{i=1}^k \frac{(n_i - \mu_i)^2}
{\hat \mu_i (1-\hat\mu_i/N_i)}.
\label{eq:wls_binom}
\end{equation}
where the variances for the binomial distribution with expectation $\mu_i$ are $\sigma_i^2 = \var[n_i] = N_i \epsilon_i (1-\epsilon_i) = \mu_i (1-\mu_i/N_i)$. Again, we replaced $\mu_i$ in the variance by the constant estimate $\hat \mu_i$ from the previous iteration. Setting the $m$ first derivatives to 0, we obtain
\begin{equation}
\frac{\partial \lsq}{\partial p_j} =
-2\sum_{i=1}^k \frac{n_i - \mu_i}{\hat\mu_i(1-\hat\mu_i/N_i)}
 \, \frac{\partial \mu_i}{\partial p_j} = 0
\label{eq:wls_binom_diff}
\end{equation}

Like in the previous case, \eq{ml_binom_diff} and \eq{wls_binom_diff} yield identical solutions in the limit $\hat \mu_i \rightarrow \mu_i$, which is approached by iterating the minimization. Again, we obtain the maximum-likelihood estimate with the IWLS fit.

Like in the previous case, the L-IWLS fit for a linear model converges faster than the ML fit, while the IWLS fit converges more slowly than the ML fit in general.

\section{Expectation of $\lsq(\phat)$ for linear models}

We compute the expectation of $\lsq$ in \eq{vls}, evaluated at the solution $\phat$ from \eq{wls_linear_solution} for linear models with $\ex[\vec y] = \vec{X} \vec{p}$, where $\vec X$ is a fixed $k \times m$ matrix, and where the measurements $\vec y$ have a known finite covariance matrix $\vec C$.
Similar proofs are found in the literature \cite{KS}.
The covariance matrix of $\phat$ is obtained by error propagation with the matrix $\vec M = (\vec X^T C^{-1} \vec X)^{-1} \vec X^T \vec C^{-1}$ and $\phat = M \vec y$ as
\begin{equation}
\vec{C}_p = \vec{M} \vec{C} \vec{M}^T = (\vec{X}^T \vec{C}^{-1} \vec{X})^{-1},
\label{eq:cp}
\end{equation}\\
where we used that $\vec{C}^{-1}$ and $(\vec{X}^T \vec C^{-1} \vec{X})^{-1}$ are symmetric matrices.

The expectation is a linear operator. Since the solution $\phat = \vec M \vec y$ is a linear function of the measurement, we have
\begin{equation}
\ex[\phat] = \vec M \ex[\vec y] = \vec M \vec X \vec p = \vec p,
\end{equation}
in other words, $\phat$ is an unbiased estimate of $\vec p$.

We expand $\lsq$ evaluated at $\phat$,
\begin{equation}
\lsq(\phat) = \vec y^T \vec C^{-1} \vec y
- \vec y^T \vec C^{-1} \vec X \phat
- \phat^T \vec X^T \vec C^{-1} \vec y
+ \phat^T \vec X^T \vec C^{-1} \vec X \phat,
\end{equation}
which simplifies with $\vec C_p^{-1} \phat = \vec X^T \vec C^{-1} \vec y$ and \eq{cp} to
\begin{equation}
\lsq(\phat) = \vec y^T \vec C^{-1} \vec y - \phat^T \vec C_p^{-1} \phat.
\end{equation}
The scalar result of a bilinear form is trivially equal to the trace of this bilinear form, and a cyclic permutation inside the trace then yields
\begin{equation}
\lsq(\phat) = \Tr(\vec C^{-1} \vec y \vec y^T) - \Tr(\vec C_p^{-1} \phat \phat^T).
\end{equation}
We compute the expectation on both sides and get, using linearity of trace and expectation,
\begin{equation}
\ex[\lsq(\phat)] = \Tr(\vec C^{-1} \ex[\vec y \vec y^T]) - \Tr(\vec C_p^{-1} \ex[\phat \phat^T]).
\end{equation}
The definition of the covariance matrix $\vec C = \ex[\vec y \vec y^T] - \ex[\vec y] \ex[\vec y]^T$ is inserted, and vice versa for $\vec C_p$. We get
\begin{equation}
\ex[\lsq(\phat)] = \Tr(\vec C^{-1} \vec C + \vec C^{-1} \ex[\vec y] \ex[\vec y]^T)
- \Tr(\vec C_p^{-1} \vec C_p + \vec C_p^{-1} \ex[\phat] \ex[\phat^T]).
\end{equation}
The trace of a matrix multiplied with its inverse is equal to the number of diagonal elements, which is $k$ in case of $\vec C$ and $m$ in case of $\vec C_p$. We use this, $\ex[\vec y] = \vec X \vec p$, $\ex[\phat] = \vec p$, and again the linearity of the trace, to get
\begin{equation}
\ex[\lsq(\phat)] = k + \Tr(\vec C^{-1} \vec X \vec p \vec p^T \vec X^T)
- \big(m + \Tr(\vec C_p^{-1} \vec p \vec p^T)\big).
\end{equation}
The remaining traces are identical and cancel,
\begin{equation}
\Tr(\vec C^{-1} \vec X \vec p \vec p^T \vec X^T)
= \Tr(\vec X^T \vec C^{-1} \vec X \vec p \vec p^T)
= \Tr(\vec C_p^{-1} \vec p \vec p^T),
\end{equation}
and so we finally obtain the result
\begin{equation}
\ex[\lsq(\phat)] = k - m,
\end{equation}
which is independent of the PDFs that describe the scatter of the measurements $\vec y$ around the expectation values $\ex[\vec y]$.

\end{document}